\documentclass[conference]{IEEEtran}

\ifCLASSINFOpdf
\else
\fi

\usepackage{lipsum}
\usepackage{amsfonts}
\usepackage{graphicx}
\usepackage{epstopdf}
\usepackage{pgfplots}

\usepackage{multicol}
\usepackage{multirow}
\usepackage{bm}

\usepackage{url}
\usepackage{cite}
\usepackage[hidelinks]{hyperref}
\hypersetup{
    colorlinks=true,
    linkcolor=black,
    filecolor=black, 
    citecolor=black,
    urlcolor=black,
}

\ifCLASSOPTIONcompsoc
    \usepackage[caption=false, font=normalsize, labelfont=sf, textfont=sf]{subfig}
\else
\usepackage[caption=false, font=footnotesize]{subfig}
\fi
\usepackage[export]{adjustbox}

\ifpdf
  \DeclareGraphicsExtensions{.eps,.pdf,.png,.jpg}
\else
  \DeclareGraphicsExtensions{.eps}
\fi

\usepackage{amsopn}
\usepackage{mathtools}

\usepackage[algo2e,ruled,vlined,linesnumbered]{algorithm2e}
\SetInd{0em}{0.5em}
\SetKwComment{Comment}{/\!\!/}{}

\usepgflibrary{shapes.geometric}
\usetikzlibrary{calc}
\pgfplotsset{compat=1.16}

\newcommand{\soa}{state-of-the-art }

\newcommand{\R}{\mathbb R}  
\newcommand{\N}{\mathcal N} 

\DeclarePairedDelimiterX{\norm}[1]{\lVert}{\rVert}{#1} 
\DeclarePairedDelimiterX{\abs}[1]{\lvert}{\rvert}{#1}
\DeclarePairedDelimiterX{\ip}[2]{\langle}{\rangle}{#1, #2}
\DeclarePairedDelimiterX{\pfrac}[2]{(}{)}{\frac{#1}{#2}}
\DeclareMathOperator{\sign}{sign}

\DeclareMathOperator{\ST}{\mathrm{ST}} 

\let\x\undefined
\newcommand{\x}{\bm{x}}
\let\y\undefined
\newcommand{\y}{\bm{y}}
\let\z\undefined
\newcommand{\z}{\bm{z}}

\let\bnu\undefined
\newcommand{\bnu}{\bm{\nu}}

\begin{document}
%
\title{Gabor is Enough: Interpretable Deep Denoising with a Gabor Synthesis Dictionary Prior}
\author{\IEEEauthorblockN{Nikola Janju\v{s}evi\'{c}, Amirhossein Khalilian-Gourtani, and Yao Wang}
\IEEEauthorblockA{New York University, Tandon School of Engineering, Brooklyn, USA\\
Email: \{npj226,akg404,yaowang\}@nyu.edu}
}
\maketitle

\begin{abstract}
Image processing neural networks, natural and artificial, have a long history with orientation-selectivity, often described mathematically as Gabor filters. Gabor-like filters have been observed in the early layers of CNN classifiers and even throughout low-level image processing networks. In this work, we take this observation to the extreme and explicitly constrain the filters of a natural-image denoising CNN to be learned 2D real Gabor filters. Surprisingly, we find that the proposed network (GDLNet) can achieve near \soa denoising performance amongst popular fully convolutional neural networks, with only a fraction of the learned parameters. We further verify that this parameterization maintains the noise-level generalization (training vs. inference mismatch) characteristics of the base network, and investigate the contribution of individual Gabor filter parameters to the performance of the denoiser. We present positive findings for the interpretation of dictionary learning networks as performing accelerated sparse-coding via the importance of untied learned scale parameters between network layers. Our network's success suggests that representations used by low-level image processing CNNs can be as simple and interpretable as Gabor filterbanks.
\end{abstract}

\IEEEpeerreviewmaketitle
\section{Introduction and Background}
In recent years, deep neural network (DNN) and convolutional neural network (CNN) architectures have claimed \soa performance on low-level image processing tasks such as denoising\cite{DnCNN,FFDNet,Zheng_2021_CVPR}. However, most DNNs are constructed as black-box function approximators whose architectures are largely derived from trial and error \cite{buhrmester2019analysis}, ignoring the theoretical signal processing backgrounds related to the task at hand. Such networks have been shown to suffer a catastrophic performance failure when a mismatch is present between the training and inference scenarios \cite{Mohan2020,janjuvsevic2021cdlnet}. More recently, there has been a growing desire for principled approach to DNN design that leads to better analysis and possible improvement of the network\cite{janjuvsevic2021cdlnet,Lecouat2020nonlocal,Simon2019,Scetbon2019,fessler_BCDNet,malezieux2021dictionary,TDV_Pock}. To this end, in \cite{janjuvsevic2021cdlnet} we introduced an architecture (CDLNet) derived from direct parameterization of an iterative algorithm solving the convolutional dictionary learning (CDL) problem. By relating the thresholds in CDLNet to the input image noise-level, this network showed near-perfect generalization at inference to noise-levels unseen during training. Notably, CDLNet achieves similar performance to other \soa fully convolutional neural networks (FCNNs)\cite{DnCNN,FFDNet} while side-stepping the use of feature-domain processing 
and limiting the representation to only analysis and synthesis convolutional dictionaries. Such directly-parameterized dictionary learning based networks \cite{janjuvsevic2021cdlnet,Simon2019,Lecouat2020Games,fessler_BCDNet} challenge the notion of low-level image processing CNNs capturing complex signal representations in hidden feature domains, and suggest that much simpler mechanisms, such as subband representations, can account for their performance.

Many dictionary learning frameworks for low-level image processing tasks, especially denoising, learn ``Gabor-like'' dictionaries\cite{janjuvsevic2021cdlnet,Simon2019,Simoncelli2001NaturalIS}. Neuroscientific investigations suggest that V1 simple cells in primary visual cortex behave like Gabor filters \cite{jones1987evaluation,marcelja1980mathematical,daugman1980two}. The early visual cortex responds maximally to lines or edges with a specific orientation which can be achieved with Gabor filter with a proper scale, a specific direction and a spatial frequency \cite{kruger2012deep}. More recently, Gabor filters have attracted attention in CNN design, from preprocessing stage \cite{yao2016gabor}, to use of Gabor filters in early CNN layer \cite{bai2019training}, to combining Gabor and learned convolutional filters \cite{luan2018gabor}. Fixed Gabor filters have also been shown to reduce the computational complexity of CNNs, however, with a significant performance reduction \cite{sarwar2017gabor}. These works overall suggest that use of Gabor filters in CNN design leads to better performance, albeit with a more complex network, or reduced complexity with degraded performance.

The goal of this work is to quantitatively explore the Gabor-like behavior of learned filters in convolutional dictionary based neural networks and pose the question ``can Gabor-like be replaced with Gabor?''. We tackle this question by replacing the filters of CDLNet \cite{janjuvsevic2021cdlnet} with parameterized 2D Gabor functions. In place of learned filters, this network learns sets of parameters that generate Gabor filters (Section \ref{sec:gabor}). Surprisingly, we find that this representation is sufficient for achieving near \soa performance in the natural-image denoising task (Section \ref{sec:exp_single}) with respect to popular FCNNs such as DnCNN \cite{DnCNN} and FFDNet \cite{FFDNet}. We verify that the near-perfect denoising generalization of CDLNet to unseen noise-levels are retained with this parameterization (Section \ref{sec:exp_generalization}). Finally, in \cite{janjuvsevic2021cdlnet} we demonstrated the importance of untying filter weights between network layers. With the proposed Gabor filterbank representation, we further explore untying of individual parameters of the Gabor function between layers and their contribution to denoising performance(Section \ref{sec:exp_params}).

\begin{figure*}[ht]
    \centering
    \includegraphics[width=0.98\linewidth]{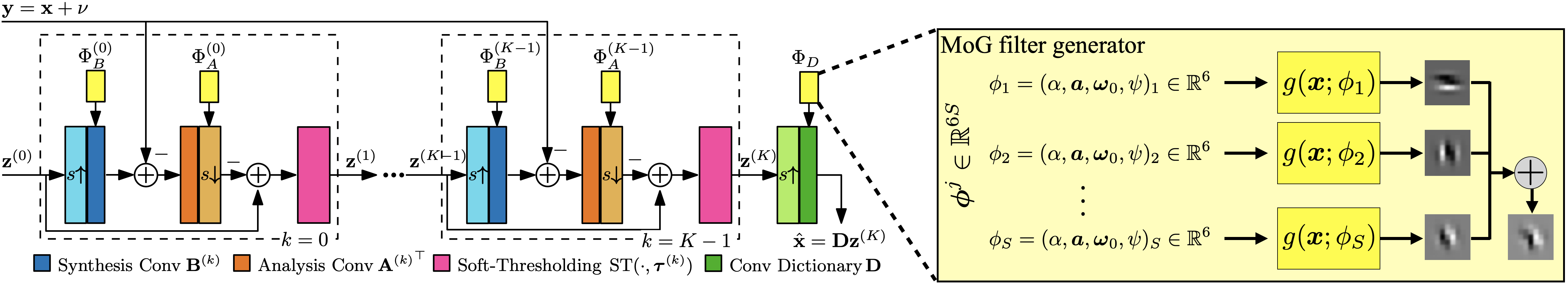}
    \caption{\textit{Left}: Block diagram of GDLNet, CDLNet with mixture of Gabor (MoG) filterbanks. Analysis and synthesis convolutions map from $1 \rightarrow M$ and $M \rightarrow 1$ channels, respectively. We emphasize that GDLNet/CDLNet do not process signals in a ``learned feature domain" in contrast to the multi-channel filtering ($M \rightarrow M$ channels) in popular DNNs such as DnCNN \cite{DnCNN}. \textit{Right}: Mixture of Gabor filters are generated as a sum of parameterized Gabor filters.}
    \vspace*{-15pt}
    \label{fig:block_diag}
\end{figure*}

\section{Problem Statement and Proposed Method}
In this manuscript, we look at the denoising of natural images via convolutional neural networks (CNNs).
We denote our observations contaminated with additive Gaussian white noise (AWGN) , $\y = \x + \bnu$, where $\bnu \sim \N(0,\sigma \boldsymbol{I})$.


\subsection{CDLNet Architecture}
The Convolutional Dictionary Learning Network (CDLNet) architecture \cite{janjuvsevic2021cdlnet} is a denoising neural network derived from the Iterative Shrinkage Thresholding Algorithm (ISTA)\cite{Beck2009}. Much like a traditional CNN, it employs layered convolutions and point-wise nonlinearities, as described below,
%
\begin{equation}
    \begin{aligned}
\hat{\x} &= \bm{D}\z^{(K)}, \quad \z^{(0)} = \bm{0}, \quad k=0,1,\dots, K-1,\\
\z^{(k+1)} &= \ST\left(\z^{(k)} - {\bm{A}^{(k)}}^T (\bm{B}^{(k)}\z^{(k)} - \y), \, \boldsymbol{\tau}^{(k)} \right). 
\end{aligned}
\label{eqn:CDLNet}
\end{equation}
Here, ${\bm{A}^{(k)}}^T, \bm{B}^{(k)}$ refer to $M$-subband analysis and synthesis filterbank convolutions, respectively, with stride-$s$, $\mathrm{ST}(\x,\tau)[n] \coloneqq \sign(\x[n])\max(0, \abs{\x[n]} - \tau)$ is known as the element-wise soft-thresholding operator, and $\bm{\tau}^{(k)} \in \R^M$ are the networks learned non-negative thresholds. The final denoised image $\hat{\x}$ is given by convolution with the network's synthesis filterbank  (dictionary), $\bm{D}$. 

CDLNet's interpretation as an accelerated sparse-coding algorithm allows for consideration of a varying input noise-level via augmenting the thresholds of the network. In \cite{janjuvsevic2021cdlnet}, it was shown that the following noise-adaptive thresholding, 
\begin{equation} \label{eqn:thresh}
\boldsymbol{\tau}^{(k)} = \boldsymbol{\tau}^{(k)}_0 + \boldsymbol{\tau}^{(k)}_1 \sigma,\quad \mathrm{such~that} \quad \boldsymbol{\tau}^{(k)}_{\{0,1\}} \geq0,
\end{equation}
with $\boldsymbol{\tau}^{(k)}_{\{0,1\}}\in\R^M$ learned parameters, yields a near perfect generalization to input noise-levels outside the network's training range at inference time. This behavior is in contrast to the catastrophic failures of black box DNNs observed outside their training ranges \cite{janjuvsevic2021cdlnet, Mohan2020}. 

\subsection{Parameterized Gabor Filterbanks} \label{sec:gabor}

The learned dictionary elements in\cite{janjuvsevic2021cdlnet} are ``Gabor-like'' filters. In this work, we consider the explicit parameterization of CDLNet's filterbanks (both analysis and synthesis) as learnable 2D real Gabor filters,
\begin{equation} \label{eq:gabor}
    g(\x; \phi) = \alpha e^{-\norm{\bm{a}\circ \x}_2^2}\cos(\boldsymbol{\omega}_0^T\x + \psi),
\end{equation}
with $\x \in \R^2$ denoting spatial position, $\circ$ denoting element-wise product, and $\phi = (\alpha, \bm{a}, \bm{\omega}_0, \psi) \in \R^6$. We refer to these as $\alpha \in \R$ the scale, $\bm{a}\in \R^2$ the (root) precision, $\bm{\omega}_0 \in \R^2$ the frequency, and $\psi \in \R$ the phase parameters. These filters are orientation selective, as seen in their Fourier domain representation of conjugate symmetric Gaussians centered at frequency $\pm\bm{\omega}_0$,
\begin{equation*}
    \mathcal{F}\{g(;\phi)\}(\bm{\omega}) = 
    \frac{\alpha}{2}\left(e^{j\psi}e^{-\norm{\frac{\bm{\omega}-\bm{\omega}_0}{2\pi\bm{a}}}_2^2} +
    e^{-j\psi}e^{-\norm{\frac{\bm{\omega}+\bm{\omega}_0}{2\pi\bm{a}}}_2^2} \right).
\end{equation*}
To further enable these filters to admit a sparse image representation, we consider parameterizing the filters of CDLNet as a \textit{mixture of Gabor} (MoG) filters,
\begin{equation}
    d(\x; \bm{\phi}) = \sum_{i=1}^S g(\x; \bm{\phi}_i).
\end{equation}
A single MoG filter of order $S > 0$ is parameterized by $\bm{\phi} \in \R^{6S}$, and a MoG $M$-subband analysis/synthesis filterbank is parameterized by $\Phi = [\bm{\phi}^1, \dots, \bm{\phi}^M] \in \R^{6SM}$. This is in contrast to the unconstrained filters in CDLNet, parameterized in $\R^{PM}$ for $\sqrt{P}\times \sqrt{P}$ filters. 

The MoG representation allows for a wider class of filters to be learned with the same computational complexity in the forward pass of the network, while maintaining the interpretablility of the Gabor parameterization over unconstrained filters. In the digital realization of these Gabor filters, we limit their spatial extent to $\sqrt{P}\times \sqrt{P}$ and evaluate \eqref{eq:gabor} at lattice points. Interestingly, the spatial extent of these filters has no effect on their parameter count. We refer to the CDLNet architecture with parameterized mixture of Gabor filters as the Gabor Dictionary Learning Network (GDLNet). Its parameters are given by $\Theta = (\{ \Phi_A^{(k)}, \Phi_B^{(k)}, \bm{\tau}^{(k)} \}_{k=0}^{K-1}, \Phi_D)$, where $\Phi_A^{(k)}, \Phi_B^{(k)}, $ refer to the Gabor parameters of the analysis, synthesis filterbanks of the $k$th layer, and similarly $\Phi_D$ refers to the Gabor parameters of the synthesis dictionary $\bm{D}$ (see \eqref{eqn:CDLNet}). Figure \ref{fig:block_diag} shows an overview of the GDLNet architecture with MoG filterbanks.

\section{Experimental Results}
\subsection{Training and Inference Setup}
We follow the training and inference setup of CDLNet \cite{janjuvsevic2021cdlnet}. All models are trained on the BSD432 dataset \cite{bsd} under supervised learning with the mean-squared-error loss,
\begin{equation} \label{eqn:mse}
\underset{
\substack{
\Phi_A^{(k)}, ~ \Phi_B^{(k)}, ~ \Phi_D,\\
\boldsymbol{\tau}^{(k)} \geq 0
}
}{\mathrm{minimize}} \quad \sum_{(\y_i,\x_i) \in \mathcal{D}} \|\x_i - \bm{D}\bm{z}^{(K)}(\y_i)\|_2^2,
\end{equation}
where $(\x_i,\y_i) \in \mathcal{D}$ represent ground-truth and observed images of the dataset. Data-augmentation, as well as additive noise, are applied online. Noise-level $\sigma \in \sigma^{\mathrm{train}}$ is sampled uniformly for each mini-batch sample. Noise-level adaptive models are given the ground truth noise-level at both training and inference, though common noise-level estimators \cite{Chang2000, Liu2013} may be applied instead at inference (see \cite{janjuvsevic2021cdlnet} for details).

\textbf{Architectures}: 
Suffixes (-S,-B) are used to denote models trained on a single noise-level (-S) or over a noise-level range \textit{without} adaptive thresholds (-B). Unless otherwise specified, table \ref{tab:arch} shows the hyperparameters of our proposed GDLNet. Code available at \cite{GithubLink}.

\begin{table}[tb]
\caption{Architectures of GDLNet models presented in the experiments section. $K$ layers, $M$ subbands, stride-$s$, filter-size $P$, MoG order $S$. Parameters are as shown unless otherwise specified. }
\centering
\resizebox{0.95\linewidth}{!}{%
\begin{tabular}{ccccccc} \hline
     Name & $K$ & $M$ & $s$ & $P$ & $S$ & Params \\ \hline
     CDLNet(-S) \cite{janjuvsevic2021cdlnet}& 30 & 169 & 2 & $7\times 7$ & n/a & 507k \\ 
     GDLNet(-S) (MoG 1) & 30 & 169 & 2 & $11\times 11$ & 1 & 66k \\ 
     GDLNet(-S,-B) (MoG 3)  & 30 & 169 & 2 & $11\times 11$ & 3 & 188k \\ \hline
\end{tabular}
}
\vspace*{-15pt}
\label{tab:arch}
\end{table}

\textbf{Initialization}: As in \cite{janjuvsevic2021cdlnet}, all filterbanks are initialized with the same parameters and subsequently normalized by the their spectral norm, $L=\norm{\bm{D}^T\bm{D}}_2$. For the Gabor filters, this normalization takes place via the scale parameter, $\alpha \gets \alpha / \sqrt{L}$.

\textbf{Gabor Filters}:
The Gabor filter parameterization requires that we recompute the filters from their parameters after each gradient update of the optimizer. This offers a minor computational overhead during training, however the filter weights may be computed and stored for faster inference once training is complete. 

\subsection{Single Noise-level Denoising} \label{sec:exp_single}
Single noise-level grayscale denoising results, as compared to \soa popular FCNNs, are shown in Table \ref{table:single}. The non-learned BM3D method \cite{bm3d} is given as baseline. GDLNet-S (MoG 1) performs competitively in the low parameter count regime, outperforming CSCNet \cite{Simon2019} and small-CDLNet-S \cite{janjuvsevic2021cdlnet}. The use of higher order MoG filters in GDLNet-S (MoG 3) allows it to perform on par with DnCNN \cite{DnCNN} and FFDNet \cite{FFDNet} while using less than half the learned parameters.

\begin{table}[ht]
\centering
\caption{Grayscale image denoising performance (PSNR) on BSD68 testset ($\sigma = \sigma^{\mathrm{train}} = \sigma^{\mathrm{test}}$). All learned models trained on BSD432\cite{bsd}. $^\dagger$ trained on CBSD432 $+$ Waterloo ED \cite{ma2017waterloo}.}
\resizebox{0.9\linewidth}{!}{%
\begin{tabular}{ccccccc} \hline
\multirow{2}{*}{Model} & \multirow{2}{*}{Params} & \multicolumn{3}{c}{Noise-level ($\sigma$)} \\
 & & 15 & 25 & 50 \\ \hline
BM3D \cite{bm3d}        &  -  & 31.07 & 28.57 & 25.62 \\
CSCNet \cite{Simon2019}$^\dagger$ & 64k & 31.57 & 29.11 & 26.24 \\
small-CDLNet-S \cite{janjuvsevic2021cdlnet} & 64k & 31.60 & 29.11 & 26.19 \\
GDLNet-S (MoG 1)        & 66k  & 31.59 & 29.13 & 26.21 \\\hline
FFDNet \cite{FFDNet}    & 485k& 31.63 & 29.19 & 26.29 \\
DnCNN \cite{DnCNN}      & 556k& \underline{31.72}& \underline{29.22} & {26.23} \\
CDLNet-S \cite{janjuvsevic2021cdlnet} & 507k& \bf{31.74} & \bf{29.26} & \bf{26.35} \\
GDLNet-S (MoG 3)        & 188k& 31.68  & \underline{29.22}  & \underline{26.30}  \\\hline
\end{tabular}
}
\label{table:single}
\end{table}

MoG order and filter-size hyperparameters used in Table \ref{table:single} are verified in Tables \ref{table:MoG} and \ref{table:ks}. An important boost in performance is given by MoG orders $>1$, though diminishing returns are observed above order 3. Minimal performance gains are observed as the filter size increases beyond $11\times 11$.

\begin{table}[ht]
\centering
\caption{Effect of mixture of Gabor (MoG) order for GDLNet-S on denoising performance (PSNR) over BSD68 testset \cite{bsd}  ($\sigma = \sigma^{\mathrm{train}} = \sigma^{\mathrm{test}}$).}
\resizebox{0.8\linewidth}{!}{%
\begin{tabular}{ccccccc} \hline
\multirow{2}{*}{MoG order} & \multirow{2}{*}{Params} & \multicolumn{3}{c}{Noise-level ($\sigma$)} \\
& & 15 & 25 & 50 \\ \hline
1 & 66k  & 31.59 & 29.13 & 26.21 \\
2 & 127k & 31.65 & 29.18 & 26.27 \\
3 & 188k & 31.68 & 29.22 & 26.30 \\
4 & 248k & 31.70 & 29.23 & 26.31 \\
5 & 309k & 31.70 & 29.24 & 26.33 \\\hline
\end{tabular}
}
\vspace*{-10pt}
\label{table:MoG}
\end{table}

\begin{table}[ht]
\centering
\caption{Effect of filter size for GDLNet-S (MoG 1) on denoising performance (PSNR) over grayscale Kodak dataset \cite{Kodak}  ($\sigma = \sigma^{\mathrm{train}} = \sigma^{\mathrm{test}}$).}
\resizebox{0.75\linewidth}{!}{%
\begin{tabular}{cccccc} \hline
\multirow{2}{*}{Filter size (P)} & \multicolumn{3}{c}{Noise-level ($\sigma$)} \\
 & 15 & 25 & 50 \\ \hline
$3\times 3$   & 32.45 & 30.06 & 27.21 \\
$7\times 7$   & 32.55 & 30.17 & 27.36 \\
$11\times 11$ & 32.57 & 30.21 & 27.40 \\
$15\times 15$ & 32.58 & 30.22 & 27.38 \\\hline
\end{tabular}
}
\vspace*{-15pt}
\label{table:ks}
\end{table}

\subsection{Blind Denoising and Generalization} \label{sec:exp_generalization}
\begin{figure}[ht] 
    \centering
    \subfloat[PSNR plot $\sigma^{\mathrm{train}}={[01,20]}$ \label{fig:GrayBlindPlot0120}]{%
        \includegraphics[width=0.49\linewidth]{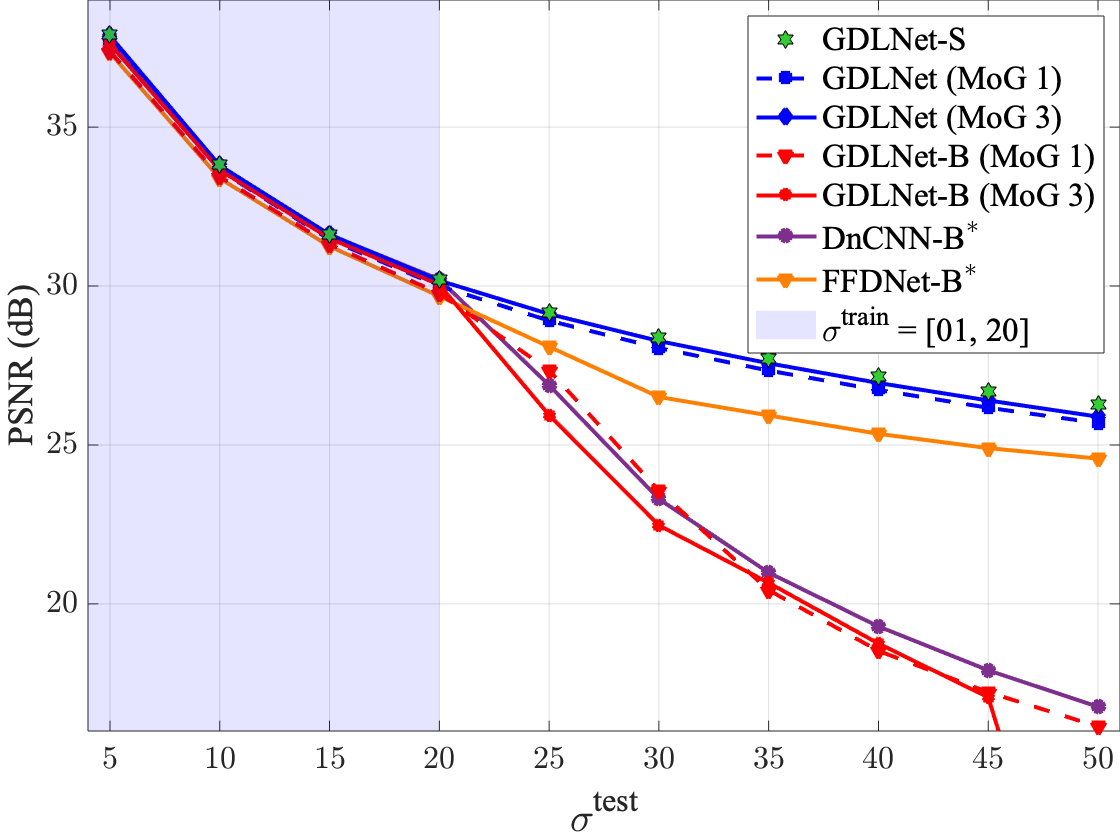}}
  \subfloat[PSNR plot $\sigma^{\mathrm{train}}={[20,30]}$ \label{fig:GrayBlindPlot2030}]{%
        \includegraphics[width=0.49\linewidth]{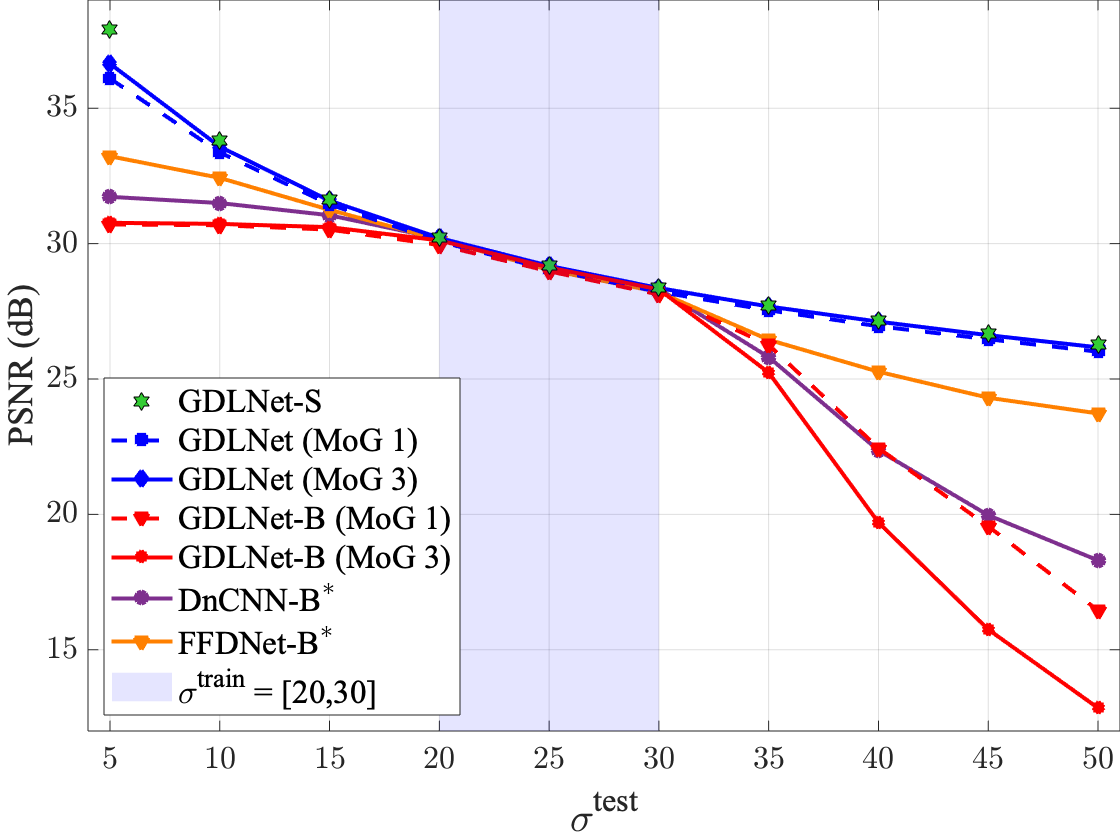}}
    \\\vspace*{-10pt}
    \subfloat[Visual comparison for $\sigma^{\mathrm{train}}={[01,20]}$ and $\sigma^{\mathrm{test}}={50}$ \label{fig:ImgGray0120}]{%
       \includegraphics[width=\linewidth]{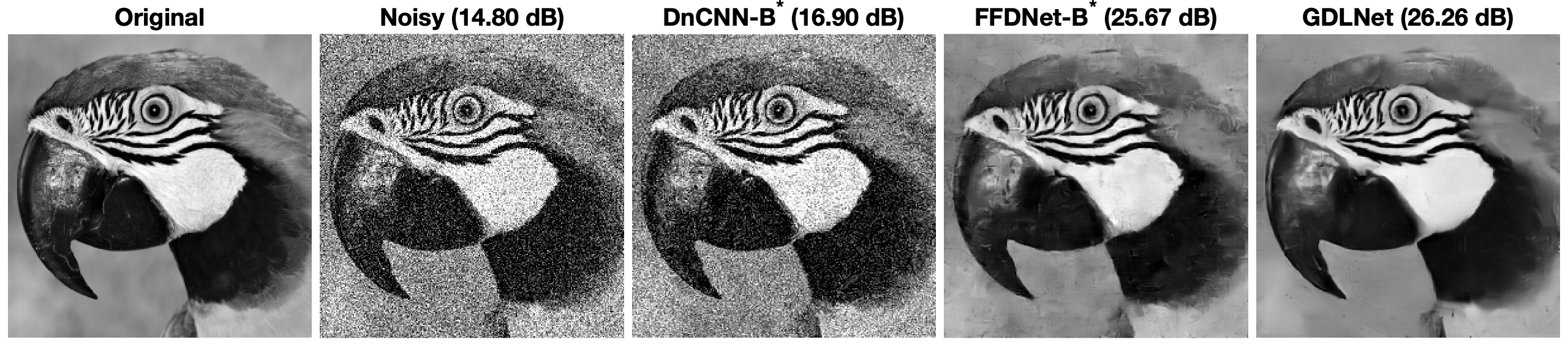}}
    \\\vspace*{-10pt}
    \subfloat[Visual comparison for $\sigma^{\mathrm{train}}={[20,30]}$ and $\sigma^{\mathrm{test}}={50}$ \label{fig:ImgGray2030} ]{%
       \includegraphics[width=\linewidth]{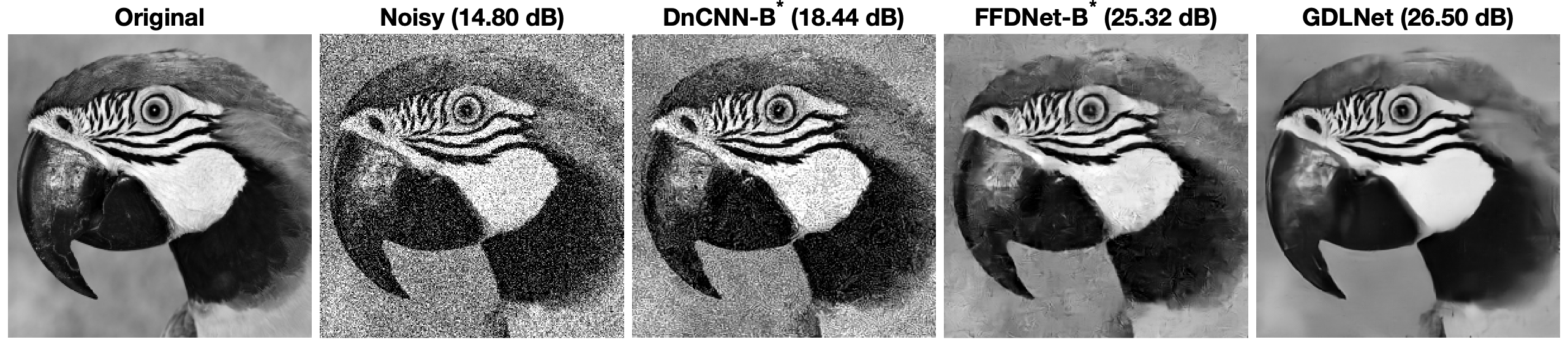}}
  \caption{(a,b) Performance of different denoising networks trained on $\sigma^{\mathrm{train}}$ and tested on $\sigma^{\mathrm{test}}$. Average PSNR calculated over BSD68 \cite{bsd}. (c,d) Visual comparison of different networks tested on noise-level $\sigma^{\mathrm{test}}=50$. GDLNet is MoG order 3. Details are better visible by zooming.}
  \vspace*{-15pt}
  \label{fig:ImgGray} 
\end{figure}

We consider the blind denoising and generalization scenarios, and compare GDLNet models of different MoG order equipped with adaptive thresholds. In Figures \ref{fig:GrayBlindPlot0120} and \ref{fig:GrayBlindPlot2030}, we show the performance of the models trained on the noise range $\sigma^{\mathrm{train}}=[1,20]$ and $[20,30]$, respectively, and tested on different noise-levels $\sigma^{\mathrm{test}}\in[5,50]$ for grayscale images. The blind denoising version of DnCNN \cite{DnCNN}, DnCNN-B, and FFDNet trained on $\sigma^{\mathrm{train}}=[01,20]$ and $[20,30]$, are as reported in \cite{janjuvsevic2021cdlnet}. The single points on these plots show the performance of GDLNet-S (MoG 3) model with single noise-level training (i.e. $\sigma^{\mathrm{train}}=\sigma^{\mathrm{test}}$).
As shown in Figures \ref{fig:GrayBlindPlot0120} and \ref{fig:GrayBlindPlot2030}, all networks perform closely over the training noise-range. Conversely, we observe the catastrophic failure of models without adaptation when generalizing outside the training noise-level. GDLNet nearly matches the performance of the models trained for a specific noise-level (GDLNet-S) across the test range. We observe a slight improved performance for GDLNet with higher order. Note that the CDLNet model architecture disentangles the noise generalization from the dictionary structure. Consequently, GDLNets of different order show similar generalization capability when equipped with adaptive thresholds. Visual comparisons in Figures \ref{fig:ImgGray0120} and \ref{fig:ImgGray2030} show superior performance of GDLNet compared to other models.

\subsection{Comparison of Learned Dictionaries}
\begin{figure}[ht]
    \centering
    \includegraphics[width=\linewidth]{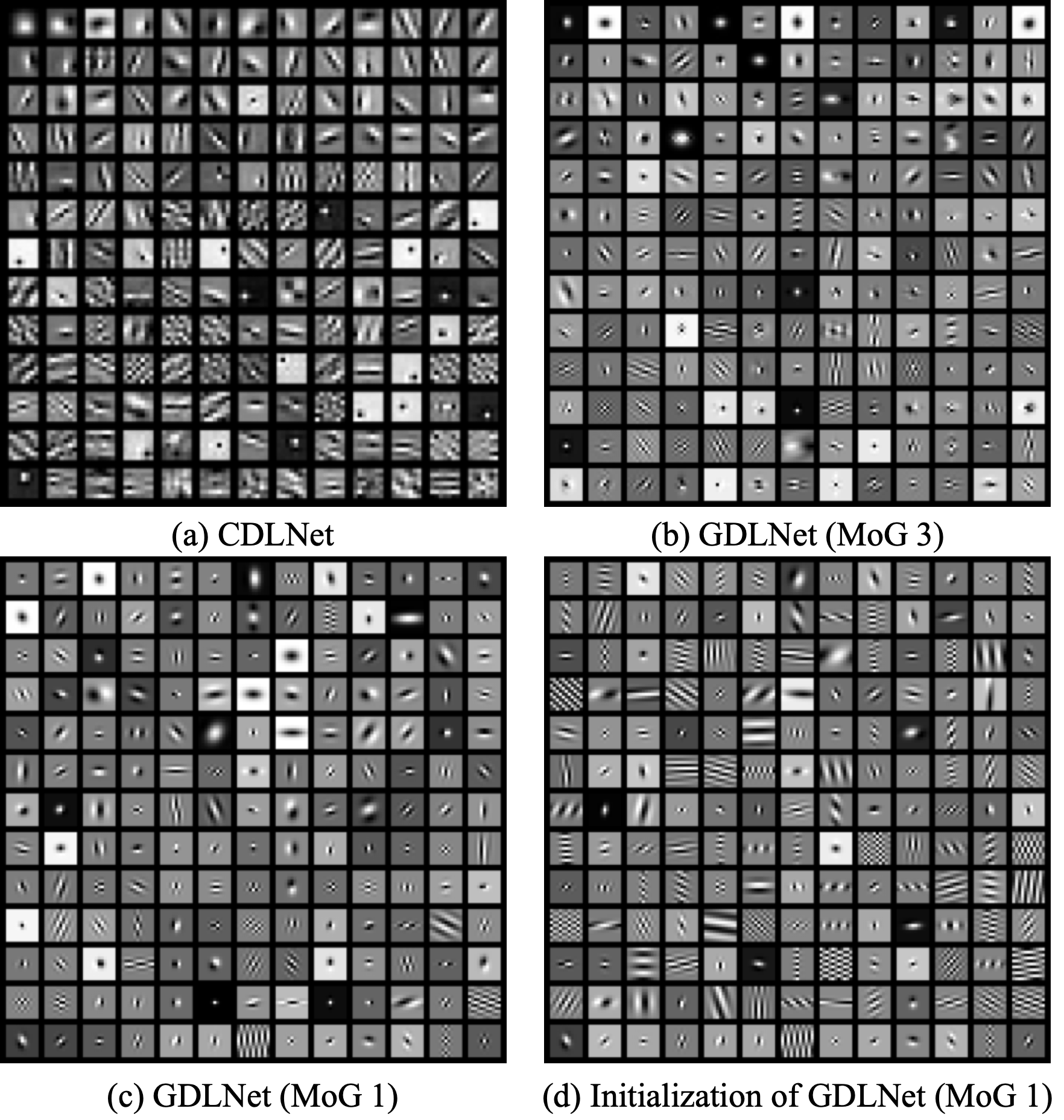}
    \caption{Synthesis dictionary ($\bm{D}$) filters for (a) CDLNet \cite{janjuvsevic2021cdlnet}, (b) GDLNet with MoG order 3, (c) GDLNet with MoG order 1, and (d) initial Gabor dictionary of (c). All models trained on $\sigma^{\mathrm{train}}=[20,30]$. Filters are ordered by relative usage over the BSD68 dataset \cite{bsd}.}
    \vspace*{-15pt}
    \label{fig:Dictionary}
\end{figure}

Figure \ref{fig:Dictionary} shows the final synthesis dictionaries of CDLNet, GDLNet, and GDLNet's initial dictionary. The use of parameterized Gabor filters adds to the interpretability of GDLNet as we can understand the subband representations simply as (a sum of) orientation features. GDLNet has cleaner looking filters and does not exhibit phase-shifted copies of filters, as in seen CDLNet Fig. \ref{fig:Dictionary}a. This is because each filter in GDLNet is centered, by construction. As shown in Fig. \ref{fig:Dictionary}a, dictionary in GDLNet MoG order 3 exhibits complex structures that are not simply a single orientation, showing that the additional flexibility of the MoG representation is contributing to the performance gain. The random initialization of Gabor parameters $\phi$ yields a good starting point for the dictionary (Fig. \ref{fig:Dictionary}d) and the network is free to learn different orientations and precisions as desired (Fig. \ref{fig:Dictionary}c). 

\subsection{Ablation Study of Gabor Filterbanks} \label{sec:exp_params}
The success of CDLNet, in contrast to CSCNet \cite{Simon2019} and others \cite{Sreter2018,Scetbon2019}, is largely attributed to untying of the weights between layers of the network. However, CDLNet and other directly parameterized unrolled networks \cite{Simon2019,Lecouat2020nonlocal,Scetbon2019,Sreter2018} are often interpreted as performing an accelerated sparse coding, either via a learned preconditioning and/or learned step-sizes of the original algorithm \cite{ablin2019learning, Chen2018}. Untying the weights between layers makes this interpretation more difficult as the network is able to vary arbitrarily from layer-to-layer. 

\begin{figure}[ht]
    \centering
    \includegraphics[width=\linewidth]{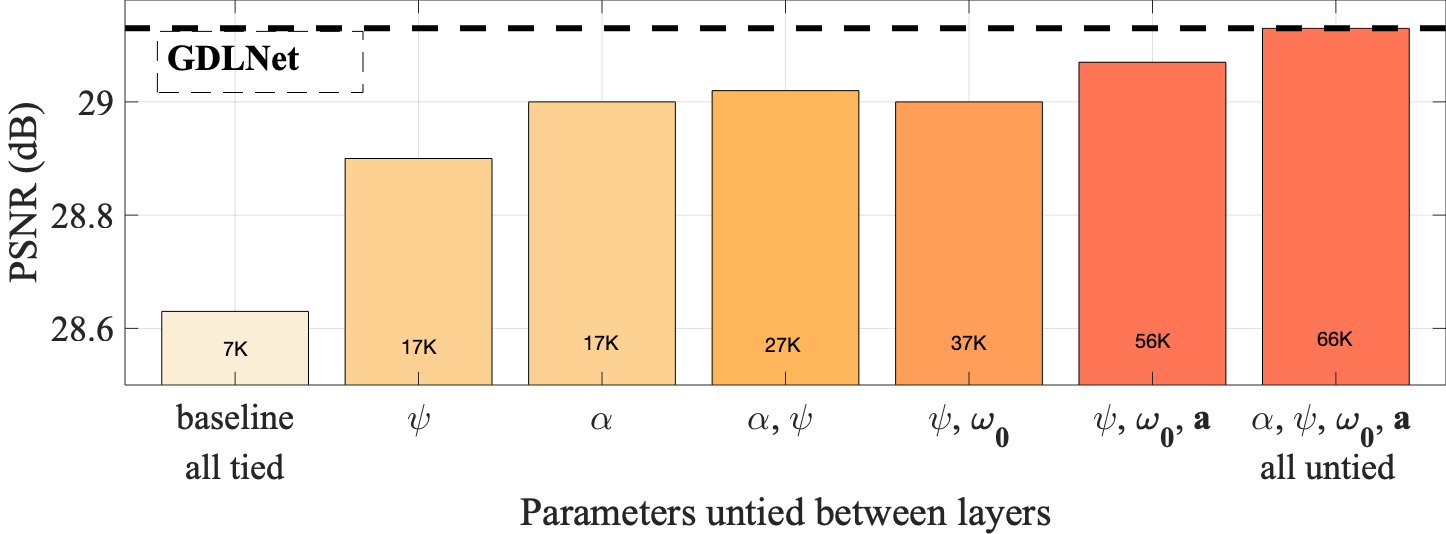}
    \caption{Effect of untied parameters between layers $k=0,1,\dots, K-1$ on performance of GDLNet (MoG 1). PSNR averaged over BSD68 \cite{bsd}, $\sigma^{\mathrm{train}} = \sigma^{\mathrm{test}} = 25$. Refer to \eqref{eq:gabor} for description of Gabor filter parameters.}
    \vspace*{-15pt}
    \label{fig:paramshare}
\end{figure}

GDLNet allows us to examine the relationship of weight tying more closely by only coupling certain parameters of the Gabor filters. Figure \ref{fig:paramshare} shows the performance of GDLNet as we progressively untie parameters between layers. While untying the phase parameter ($\psi$) shows increased performance compared to baseline, very significant increase is seen by simply untying the scale parameter ($\alpha$). To the benefit of the sparse-coding interpretation of the network, the scale parameter ($\alpha$) corresponds most closely to having learned step-sizes within the original algorithm. The minimal performance improvement observed when phase and scale are untied further verifies the sparse-coding interpretation of the network. In contrast, untying the phase and frequency alone does not yield as great of an improvement. This can be justified by considering that phase and frequency parameters may have a considerable effect on the overall magnitude of the filter, and thus can indirectly effect the implicit learned step-sizes.

\section{Conclusion}
In this work, we replaced the filters of an unrolled dictionary learning network (CDLNet) with parameterized 2D real Gabor functions. By using a mixture of Gabor filters, we achieved near \soa denoising performance at a fraction of parameter count and retained the network's generalization behavior in unseen noise-levels. Progressively untying the Gabor filter parameters allowed us to understand the contribution of each component to denoising performance and strengthened the sparse-coding interpretation of our network. In future, we aim to further investigate the use of Gabor representation in imaging inverse problems. The simplicity of this representation is suited for a better understanding of the network and may also be exploited for novel capabilities.



\bibliographystyle{IEEEtran}
\bibliography{refs}

\begin{thebibliography}{10}
\providecommand{\url}[1]{#1}
\csname url@samestyle\endcsname
\providecommand{\newblock}{\relax}
\providecommand{\bibinfo}[2]{#2}
\providecommand{\BIBentrySTDinterwordspacing}{\spaceskip=0pt\relax}
\providecommand{\BIBentryALTinterwordstretchfactor}{4}
\providecommand{\BIBentryALTinterwordspacing}{\spaceskip=\fontdimen2\font plus
\BIBentryALTinterwordstretchfactor\fontdimen3\font minus
  \fontdimen4\font\relax}
\providecommand{\BIBforeignlanguage}[2]{{%
\expandafter\ifx\csname l@#1\endcsname\relax
\typeout{** WARNING: IEEEtran.bst: No hyphenation pattern has been}%
\typeout{** loaded for the language `#1'. Using the pattern for}%
\typeout{** the default language instead.}%
\else
\language=\csname l@#1\endcsname
\fi
#2}}
\providecommand{\BIBdecl}{\relax}
\BIBdecl

\bibitem{DnCNN}
K.~Zhang, W.~Zuo, Y.~Chen, D.~Meng, and L.~Zhang, ``Beyond a gaussian denoiser:
  Residual learning of deep {CNN} for image denoising,'' \emph{IEEE
  Transactions on Image Processing}, vol.~26, no.~7, p. 3142–3155, 2017.

\bibitem{FFDNet}
K.~Zhang, W.~Zuo, and L.~Zhang, ``{FFDNet}: Toward a fast and flexible solution
  for {CNN}-based image denoising,'' \emph{IEEE Transactions on Image
  Processing}, vol.~27, no.~9, p. 4608–4622, 2018.

\bibitem{Zheng_2021_CVPR}
H.~Zheng, H.~Yong, and L.~Zhang, ``Deep convolutional dictionary learning for
  image denoising,'' in \emph{Proceedings of the IEEE/CVF Conference on
  Computer Vision and Pattern Recognition (CVPR)}, June 2021, pp. 630--641.

\bibitem{buhrmester2019analysis}
V.~Buhrmester, D.~M{\"u}nch, and M.~Arens, ``Analysis of explainers of black
  box deep neural networks for computer vision: A survey,'' \emph{arXiv
  preprint arXiv:1911.12116}, 2019.

\bibitem{Mohan2020}
S.~Mohan, Z.~Kadkhodaie, E.~P. Simoncelli, and C.~Fernandez-Granda, ``Robust
  and interpretable blind image denoising via bias-free convolutional neural
  networks,'' in \emph{International Conference on Learning Representations
  (ICLR)}, 2020.

\bibitem{janjuvsevic2021cdlnet}
N.~Janju{\v{s}}evi{\'c}, A.~Khalilian-Gourtani, and Y.~Wang, ``{CDLNet}:
  Noise-adaptive convolutional dictionary learning network for blind denoising
  and demosaicing,'' \emph{arXiv preprint arXiv:2112.00913}, 2021.

\bibitem{Lecouat2020nonlocal}
B.~Lecouat, J.~Ponce, and J.~Mairal, ``Fully trainable and interpretable
  non-local sparse models for image restoration,'' in \emph{European Conference
  on Computer Vision (ECCV)}, 2020.

\bibitem{Simon2019}
D.~Simon and M.~Elad, ``Rethinking the {CSC} model for natural images,'' in
  \emph{Advances in Neural Information Processing Systems}, 2019, pp.
  2274--2284.

\bibitem{Scetbon2019}
M.~Scetbon, M.~Elad, and P.~Milanfar, ``Deep k-svd denoising,'' \emph{ArXiv},
  vol. abs/1909.13164, 2019.

\bibitem{fessler_BCDNet}
S.~Ravishankar, I.~Y. Chun, and J.~A. Fessler, ``Physics-driven deep training
  of dictionary-based algorithms for mr image reconstruction,'' in \emph{2017
  51st Asilomar Conference on Signals, Systems, and Computers}, 2017, pp.
  1859--1863.

\bibitem{malezieux2021dictionary}
B.~Mal{\'e}zieux, T.~Moreau, and M.~Kowalski, ``Dictionary and prior learning
  with unrolled algorithms for unsupervised inverse problems,'' \emph{arXiv
  preprint arXiv:2106.06338}, 2021.

\bibitem{TDV_Pock}
E.~Kobler, A.~Effland, K.~Kunisch, and T.~Pock, ``Total deep variation for
  linear inverse problems,'' in \emph{IEEE Conference on Computer Vision and
  Pattern Recognition}, 2020.

\bibitem{Lecouat2020Games}
B.~Lecouat, J.~Ponce, and J.~Mairal, ``A flexible framework for designing
  trainable priors with adaptive smoothing and game encoding,'' in
  \emph{Advances in Neural Information Processing Systems}, vol.~33, 2020, pp.
  15\,664--15\,675.

\bibitem{Simoncelli2001NaturalIS}
E.~P. Simoncelli and B.~A. Olshausen, ``Natural image statistics and neural
  representation.'' \emph{Annual review of neuroscience}, vol.~24, pp.
  1193--216, 2001.

\bibitem{jones1987evaluation}
J.~P. Jones and L.~A. Palmer, ``An evaluation of the two-dimensional gabor
  filter model of simple receptive fields in cat striate cortex,''
  \emph{Journal of neurophysiology}, vol.~58, no.~6, pp. 1233--1258, 1987.

\bibitem{marcelja1980mathematical}
S.~Mar{\^c}elja, ``Mathematical description of the responses of simple cortical
  cells,'' \emph{JOSA}, vol.~70, no.~11, pp. 1297--1300, 1980.

\bibitem{daugman1980two}
J.~G. Daugman, ``Two-dimensional spectral analysis of cortical receptive field
  profiles,'' \emph{Vision research}, vol.~20, no.~10, pp. 847--856, 1980.

\bibitem{kruger2012deep}
N.~Kruger, P.~Janssen, S.~Kalkan, M.~Lappe, A.~Leonardis, J.~Piater, A.~J.
  Rodriguez-Sanchez, and L.~Wiskott, ``Deep hierarchies in the primate visual
  cortex: What can we learn for computer vision?'' \emph{IEEE transactions on
  pattern analysis and machine intelligence}, vol.~35, no.~8, pp. 1847--1871,
  2012.

\bibitem{yao2016gabor}
H.~Yao, L.~Chuyi, H.~Dan, and Y.~Weiyu, ``Gabor feature based convolutional
  neural network for object recognition in natural scene,'' in \emph{2016 3rd
  International Conference on Information Science and Control Engineering
  (ICISCE)}.\hskip 1em plus 0.5em minus 0.4em\relax IEEE, 2016, pp. 386--390.

\bibitem{bai2019training}
J.~Bai, Y.~Zeng, Y.~Zhao, and F.~Zhao, ``Training a v1 like layer using gabor
  filters in convolutional neural networks,'' in \emph{2019 International Joint
  Conference on Neural Networks (IJCNN)}.\hskip 1em plus 0.5em minus
  0.4em\relax IEEE, 2019, pp. 1--8.

\bibitem{luan2018gabor}
S.~Luan, C.~Chen, B.~Zhang, J.~Han, and J.~Liu, ``Gabor convolutional
  networks,'' \emph{IEEE Transactions on Image Processing}, vol.~27, no.~9, pp.
  4357--4366, 2018.

\bibitem{sarwar2017gabor}
S.~S. Sarwar, P.~Panda, and K.~Roy, ``Gabor filter assisted energy efficient
  fast learning convolutional neural networks,'' in \emph{2017 IEEE/ACM
  International Symposium on Low Power Electronics and Design (ISLPED)}.\hskip
  1em plus 0.5em minus 0.4em\relax IEEE, 2017, pp. 1--6.

\bibitem{Beck2009}
A.~Beck and M.~Teboulle, ``A fast iterative shrinkage-thresholding algorithm
  for linear inverse problems,'' \emph{SIAM Journal on Imaging Sciences},
  vol.~2, pp. 183--202, 01 2009.

\bibitem{bsd}
D.~Martin, C.~Fowlkes, D.~Tal, and J.~Malik, ``A database of human segmented
  natural images and its application to evaluating segmentation algorithms and
  measuring ecological statistics,'' in \emph{Proceedings of 8th IEEE
  International Conference on Computer Vision (ICCV)}, vol.~2, 2001, pp.
  416--423.

\bibitem{Chang2000}
S.~G. Chang, B.~Yu, and M.~Vetterli, ``Adaptive wavelet thresholding for image
  denoising and compression,'' \emph{{IEEE} Transactions on Image Processing},
  vol.~9, no.~9, pp. 1532--1546, 2000.

\bibitem{Liu2013}
X.~Liu, M.~Tanaka, and M.~Okutomi, ``Single-image noise level estimation for
  blind denoising,'' \emph{{IEEE} Transactions on Image Processing}, vol.~22,
  no.~12, pp. 5226--5237, Dec. 2013.

\bibitem{GithubLink}
\BIBentryALTinterwordspacing
N.~Janju\v{s}evi\'{c}, \emph{CDLNet repository}, 2021. [Online]. Available:
  \url{https://github.com/nikopj/CDLNet-OJSP}
\BIBentrySTDinterwordspacing

\bibitem{bm3d}
K.~Dabov, A.~Foi, V.~Katkovnik, and K.~Egiazarian, ``Image denoising by sparse
  3-{D} transform-domain collaborative filtering,'' \emph{IEEE Transactions on
  Image Processing}, vol.~16, no.~8, pp. 2080--2095, 2007.

\bibitem{ma2017waterloo}
K.~Ma, Z.~Duanmu, Q.~Wu, Z.~Wang, H.~Yong, H.~Li, and L.~Zhang, ``{Waterloo
  Exploration Database}: New challenges for image quality assessment models,''
  \emph{IEEE Transactions on Image Processing}, vol.~26, no.~2, pp. 1004--1016,
  Feb. 2017.

\bibitem{Kodak}
\BIBentryALTinterwordspacing
R.~Franzen, \emph{The Kodak Color Image Dataset,}. [Online]. Available:
  \url{http://r0k.us/graphics/kodak/}
\BIBentrySTDinterwordspacing

\bibitem{Sreter2018}
H.~Sreter and R.~Giryes, ``Learned convolutional sparse coding,'' in
  \emph{Proceedings of IEEE International Conference on Acoustics, Speech and
  Signal Processing (ICASSP)}, 2018, pp. 2191--2195.

\bibitem{ablin2019learning}
P.~Ablin, T.~Moreau, M.~Massias, and A.~Gramfort, ``Learning step sizes for
  unfolded sparse coding,'' in \emph{Advances in Neural Information Processing
  Systems}, 2019, pp. 13\,100--13\,110.

\bibitem{Chen2018}
X.~Chen, J.~Liu, Z.~Wang, and W.~Yin, ``Theoretical linear convergence of
  unfolded {ISTA} and its practical weights and thresholds,'' in \emph{Advances
  in Neural Information Processing Systems}, vol.~31, 2018.

\end{thebibliography}
%
%
%

\end{document}